\documentclass[runningheads]{llncs}
\usepackage{graphicx}
\usepackage{hyperref}
\usepackage{cleveref}
\usepackage[table,xcdraw]{xcolor}
\usepackage{booktabs} % For formal tables
\usepackage{diagbox}
\usepackage{multirow}
\usepackage{amssymb}
\usepackage{caption}
\usepackage{subcaption}

\begin{document}
\title{Feedback Clustering for Online Travel Agencies Searches: a Case Study}
 \author{Sara Scaramuccia\inst{1,2}
% \orcidID{0000-0003-4010-6507}
 \\
 % \and
 Simon Nanty\inst{2}
 % \orcidID{1111-2222-3333-4444}
 \and
 Florent Masseglia\inst{3}
 % \orcidID{2222--3333-4444-5555}
 }
% %
% % \authorrunning{F. Author et al.}
 \institute{Universit{\'e} de la C{\^o}te d'Azur, France 
 % \and Springer Heidelberg, Tiergartenstr. 17, 69121 Heidelberg, Germany
 \email{sara.scaramuccia@amadeus.com}
 \and {Amadeus S.A.S., Sophia Antipolis, France}
 \email{Simon.nanty@amadeus.com}
 \and {}
 \email{florent.masseglia@inria.fr}
 }
% \url{http://www.springer.com/gp/computer-science/lncs} 
% \and ABC Institute, Rupert-Karls-University Heidelberg, Heidelberg, Germany\\
\maketitle              % typeset the header of the contribution
\begin{abstract}
%%%% abstract submitted
Understanding choices performed by online customers is a growing need in the travel industry. 
In many practical situations, the only available information is the flight search query performed by the customer with no additional profile knowledge. In general, customer flight bookings are driven by prices, duration, number of connections, and so on. However, not all customers might assign the same importance to each of those criteria. 
Here comes the need of grouping together all flight searches performed by the same kind of customer, that is having the same booking criteria. Better recommendations can be proposed to customers with similar booking criteria. The effectiveness of some set of recommendations, for a single cluster, can be measured in terms of the number of bookings historically performed. This effectiveness measure plays the role of a feedback, that is an external knowledge which can be recombined to iteratively obtain a final segmentation. 
In this paper, we describe our Online Travel Agencies (OTA) flight search use case and highlight its specific features. We address the flight search segmentation problem motivated above by proposing a novel algorithm called Split-or-Merge (S/M). This algorithm is a variation of the Split-Merge-Evolve (SME) method. 
The SME method has already been introduced in the community as an iterative process updating a clustering given by the K-means algorithm by splitting and merging clusters subject to feedback independent evaluations. No previous application of the SME method to the real-word data is reported in literature to the best of our knowledge. Here, we provide experimental evaluations over real-world data to the SME and the S/M methods. The impact on our domain-specific metrics obtained under the SME and the S/M methods suggests that feedback clustering techniques can be very promising in the handling of the domain of OTA flight searches.
\end{abstract}

\section{Introduction}\label{sec:intro}
% general motivations: increasing interest in travel industry towards customer choice understanding
In the travel industry, there is a strong need in understanding customer needs for applications such as, pricing, revenue management, service development, and the one addressed in this paper, namely flight search recommendations.
Most of times customers’ interests are only expressed as a flight search request and this makes the case of flight recommendations so peculiar.
As pointed out in~\cite{Mottini+2018}, the challenge in this domain is the lack of a customer profile knowledge to rely on.
Authors in~\cite{Mottini+2018} exploit Discrete Choice Modeling to better understand customers' behaviors.
However, this implies to have some predefined customer classes.
% In~\cite{Mottini+2018}, authors tackle the problem of providing the best flight recommendations by combining Discrete Choice Modeling techniques to flight search segmentation to better understand customers' behaviors.
%

We tackle a similar task by focusing on clustering techniques. Our approach aims at  grouping together those flight searches which prize similar criteria when booking/choosing a flight, that is similar priorities assigned to recommendation features such as price, duration, number of connections, and so on.
We do not want simply to find the cluster of flight searches corresponding to some given priorities in the booking criterion.
Indeed, that would be similar to labeling a cluster as business or leisure in advance as in the case of Discrete Choice Modeling approaches.
Moreover, the similarity among flight searches in the same cluster is not necessarily correlated to some standard clustering quality measure, such as Silhouette or Adjusted Rand indexes.
Hence, we want the quality of the flight recommendations obtained for a single cluster to be treated as an external knowledge to be added to drive the clustering process.
This motivates our approach. We choose an Active Learning strategy~\cite{Settles2009}, here applied in particular to domain segmentation and called Feedback Clustering.
% finding a flight search segmentation where each segment corresponds to a kind of customer in order to maximize some recommendation feedback quantities.
% In doing that, differently from the previous works, we assign no prior meaning to each segment.

In Feedback Clustering, one starts with an initial clustering.
Then, the clustering is evaluated, that is a feedback is collected. Finally, based on that feedback the clustering model is updated and a new clustering is produced.
The iterative process stops when a threshold quality value is satisfied.
\begin{figure}
\centering
\begin{subfigure}{.2\textwidth}
  \centering
  \includegraphics[width=\linewidth]{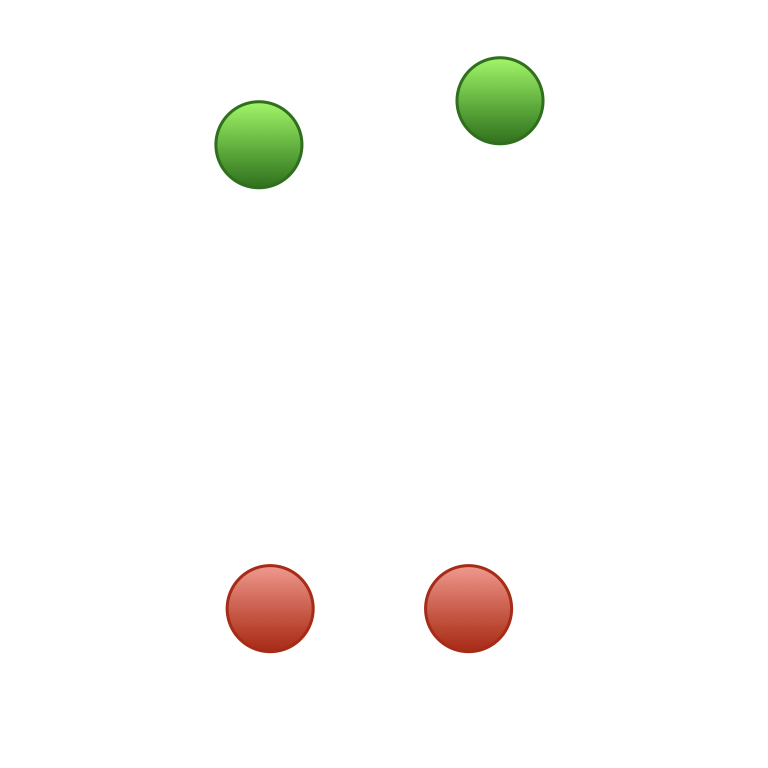}
  \caption{}
  \label{fig:ex-001}
\end{subfigure}%
\begin{subfigure}{.2\textwidth}
  \centering
  \includegraphics[width=\linewidth]{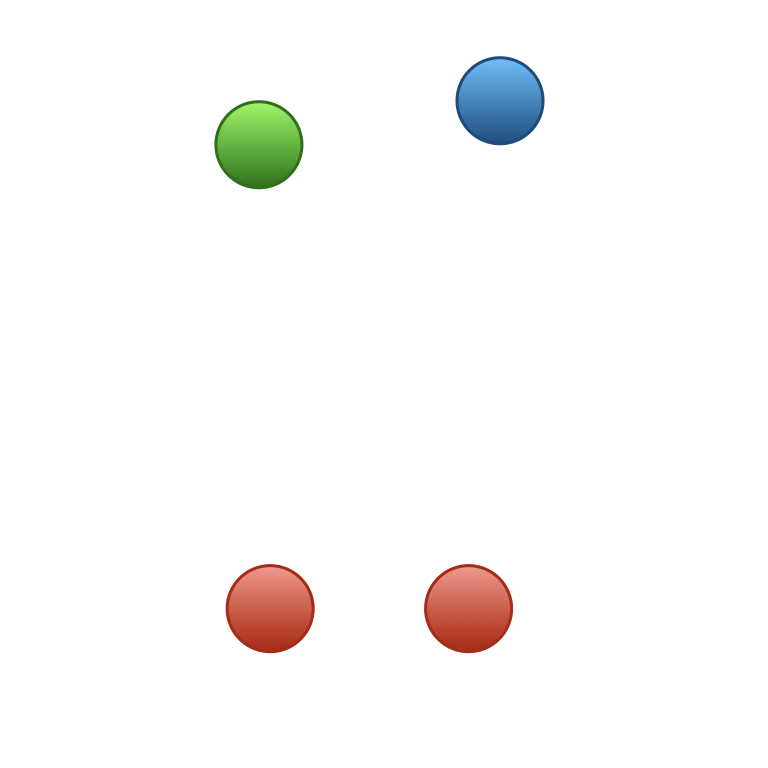}
  \caption{}
  \label{fig:ex-002}
\end{subfigure}
\begin{subfigure}{.2\textwidth}
  \centering
  \includegraphics[width=\linewidth]{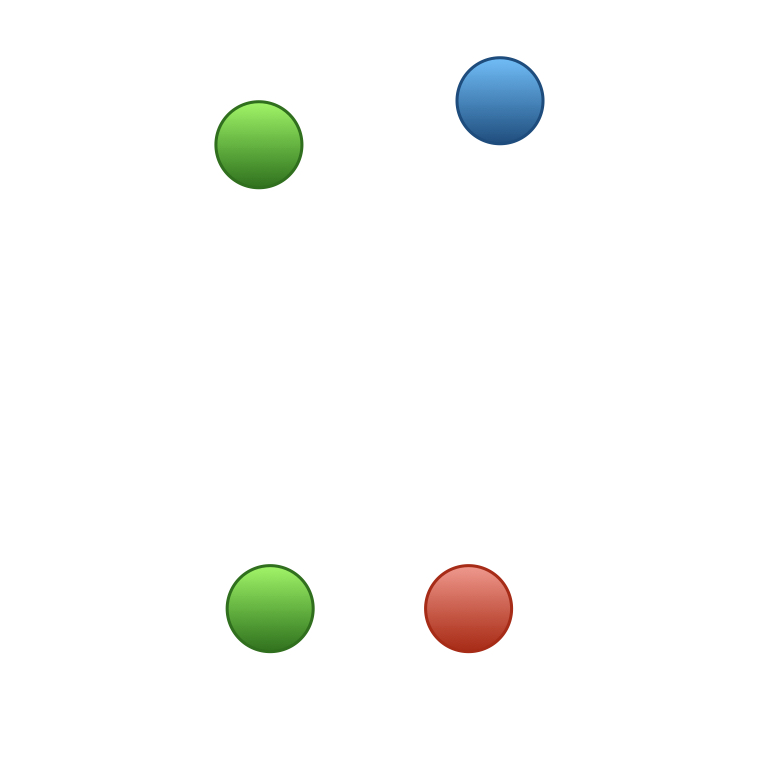}
  \caption{}
  \label{fig:ex-003}
\end{subfigure}%
\begin{subfigure}{.2\textwidth}
  \centering
  \includegraphics[width=\linewidth]{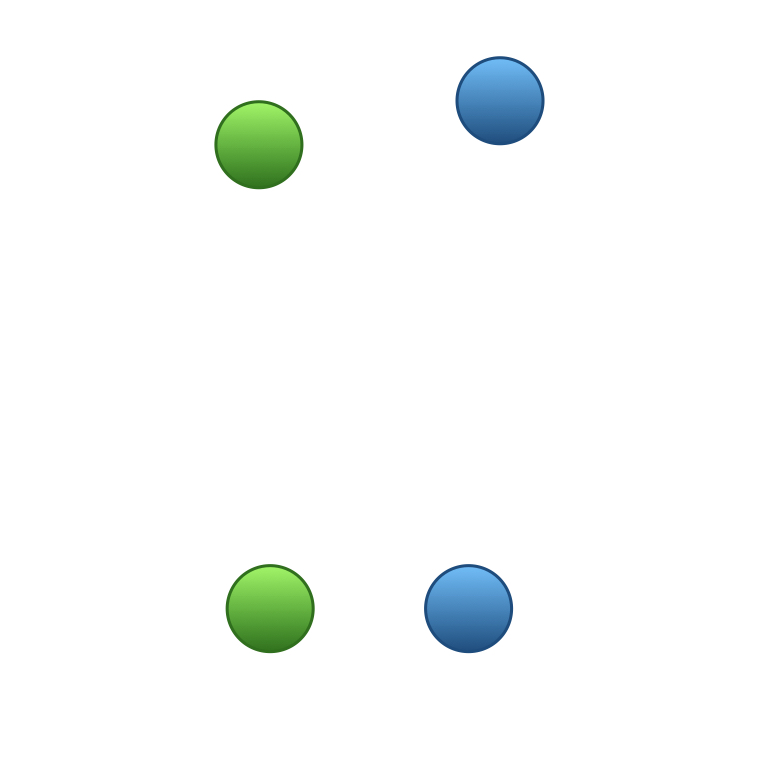}
  \caption{}
  \label{fig:ex-004}
\end{subfigure}%
\caption{Example: a segmentation where clusters (marked by colors) are progressively arranged according to the external feedback of grouping points with similar horizontal component.}
\label{fig:example}
\end{figure}
In the example in \Cref{fig:example}, we show feedback clustering in action in a toy example.
We have an initial clustering (\ref{fig:ex-001}) corresponding to the output of any clustering algorithm simply based on the geometrical arrangement of our data.
We imagine that our feedback prizes clusters where points share similar horizontal values. 
We imagine that as something we could not know in advance and thus could not have been exploited in the clustering algorithm. Rather, we think of itas a very simplified possible result of some domain analysis. In the example, the feedback is in the form of rejecting the cluster where points are to different.
Then in step (\ref{fig:ex-002}), the worst cluster is modified by splitting it into two.
Afterwards in step (\ref{fig:ex-003}), the two singletons are merged to form a cluster whose points have very similar horizontal coordinates.
Finally in step (\ref{fig:ex-004}), the final clustering is produced by merging the two left points.
The so obtained clustering reflects better than the initial one the feedback request.

In our application, we have a specific feedback we call {\em customizability}.
Customizability measures in each cluster how effective are the customer preferences estimated by an Amadeus optimization routine based on historical data.
In some sense, we are measuring how ``learnable'' the class of customer is from a  single cluster.

Since the customizability feedback is a numerical value independently associated to each cluster, we focus on the Feedback Clustering technique introduced in~\cite{Wang2018}, here called Split-Merge-Evolve (SME) method, along with specifically tailored variations.
The SME method  is an example of Feedback Clustering where the clustering model is updated from one iteration to another by independently acting over the worst found clusters which can either be split or merged to others to form better clusters.
Moreover, the SME method is presented as a modification of the $k$-means algorithm but the approach can be easily adapted to any clustering algorithm in the future.

Our main contributions are:
\begin{itemize}
 \item a discussion on the results we obtained via the SME method under both a standard cluster quality evaluation, namely the Residual Square Sum (RSS) and the non-deterministic customizability feedback 
 \item a similar discussion on a variation of the SME method here introduced and called the S/M method where split and merge operators might be applied independently, that is not necessarily in a row
 \item a comparison between the SME and S/M methods in terms of the obtained results and the future perspectives.
\end{itemize}

%structure
The paper is organized as follows.
In~\Cref{sec:rel}, we review the related works.
In~\Cref{sec:back}, we formalize our problem to be used in~\Cref{sec:back-sme} to present the details of the SME method and in~\Cref{sec:back-s/m} its variation: the S/M method.
In~\Cref{sec:exp}, we present our results and highlight our contribution.
In \Cref{sec:conclusions}, we discuss the relevance of the work and future directions.

% Moreover, .
% The SME method depends on how cluster quality is evalutated. We test it both under  .
% % we discuss strong and weak points
% We discuss differences and strong/weak points.
% 
% % we present  a version of Wang modified for our specific use case: SoM method
% Moreover, we present the same kind of results obtained under a small variation of the SME method.
% Therein, split and merge operators might be applied independently that is not necessarily in pairs. 
% 
% % we compare and discuss SME to SoM methods 
% Finally, we compare and discuss the two approaches in terms of the obtained results and the future persectives.

\section{Related Works}\label{sec:rel}
In this section, we present an overview on the related works.

% % ...and method to provide segmentations for personalized recommendations for travel searches: they are not competitors but they can be combined
% As already pointed out in~\Cref{sec:intro}, the problem of classifying customers in travel industry by simply relying on flight searches has been already set into discrete customer choice modeling terms~\cite{Mottini+2018}.
% 
% In the travel industry, there is a strong need in understanding customer needs for applications such as, pricing, revenue management, service development, and the one addressed in this paper, namely flight search recommendations.
% The challenge in this domain is the lack of a customer profile knowledge to rely on.
% In~\cite{Mottini+2018} authors exploit discrete customer choice modeling to classify flight searches among predefined classes, e.g. business and leisure customers.
%

The problem of classifying customers in the travel industry by simply relying on flight searches and no prior available classes is addressed in~\cite{SujoySimon2020arXiv}.
Therein, the segmentation is achieved by exploiting consensus clustering techniques surveyed in~\cite{Vega-Pons+2011}.
% Instead, our aim is that of finding a domain segmentation with no prior meaning associated to each segment.
% Similarly in~\cite{Sujoy}, the segmentation is achieved by exploiting consensus clustering techniques surveyed in~\cite{Vega-Pons+2011}.
Therein, consensus clustering is integrated with a multi-objective optimization process to evaluate the effectiveness of the recommendations over each segment independently.
In consensus clustering, several, heterogeneous enough, segmentation models are taken as an initial population, possibly with a different number of clusters. Afterwards, the genetic algorithm NSGA-II~\cite{Deb+2002} is adapted to find the segmentation which maximizes the consensus while minimazing the standard deviation in similarity.
% The average similarity of consensus is maximized while minimizing the standard deviation among similarity values so that to avoid to obtain a solution too close to an initial one.

% Motivated by the need of relaxing our expectations on the side of clustering quality itself while gaining flexibility in learning from the evaluation performed over each cluster for the effectiveness of the recommendations, we address in this work a similar task in a new framework.

As already explained in~\Cref{sec:intro}, our framework is a specific area of the more general framework of active learning, namely that of {\em feedback clustering}. % We refer the reader to~\cite{Settles2009} for a theoretical overview in active learning.
To the best of our knowledge there is no attempt to achieve a similar task in the framework of feedback clustering. 
%
% In feedback clustering, one starts with an initial clustering.
% Then, the clustering is evaluated, that is a feedback is collected. Finally, based on that feedback the clustering model is updated and a new clustering is produced.
%
It is worth noticing that a feedback might come in several forms (categorical/numerical) and levels (instance, cluster, clustering). Therefore, there are several and very different ways to perform feedback clustering in the literature. 

In~\cite{Caruana2006}, the authors consider {\em multiple clusterings} to be successively selected based on the user feedback with no particular restriction on the kind of feedback.

In the case of a feedback at {\em instance level}, the methods described  in~\cite{Balcan+2008}\cite{Eriksson+2011} exploit pair-wise similarity (numerical) while~\cite{Basu+2004}\cite{Klein+2002}\cite{Jain2006} exploit must/cannot link information (categorical),  among domain points. The methods in~\cite{DasguptaNg2010}\cite{WangDavidson2010} require  a preferred component direction in the feature space to be specified and exploit spectral clustering.

In the case of a feedback at {\em cluster level}, we mention~\cite{Quan+2018} for numerical (dis)similarity among clusters.
% %
% Methods in~\cite{} exploit feedback to modify the target function leading the clustering phase. Hence, this restricts the possible clustering algorithms to those supporting a target function.
%
Among the methods exploiting categorical cluster feedbacks, let us mention the work of~\cite{BalcanBlum2008}\cite{Awasthi+2013}\cite{Srivastava2016}\cite{Wang2018} based on {\em cluster rejection}.
One or more clusters are processed and the feedback consists in keeping the cluster, modifying it, or rejecting it.
Those methods are the best suited to our purposes because they act at cluster level.
In~\cite{BalcanBlum2008}, the user can reject a cluster to be split into two clusters of equal cardinality or reject two clusters to be merged. A local version of a similar approach is proposed in~\cite{Awasthi+2013}.
In~\cite{Srivastava2016}, the authors introduce a Bayesian elicitation process to let the model learn how to clusterize data based on previous feedbacks of that kind.
In~\cite{Wang2018} authors introduce the SME method by applying split/merge operators to modify rejected clusters obtained by applying $k$-means.
More precisely, the rejected cluster is, at a first step, split into two new clusters, at a second step, the two closest clusters are merged into a single one.
The method combines advantages of bisecting $k$-means clustering (the split phase) to those of agglomerative hierarchical clustering (the merge phase).
Under the Adjusted Rand Index (ARI), the SME method show its effectiveness compared to $k$-means and agglomerative hierarchical clustering over both synthetic and real datasets.

% We apply to our use case the algorithm presented in~\cite{Wang2018} along with specifically tayolored variations.
% Our choice is motivated hereafter.
% Firstly, the proposed framework is compatible with our case, that is numerical feedbacks independently given over each cluster.
% Secondly, it is presented for $k$-means algorithm but the approach can be easily adapted to any clustering algorithm in the future.
% Lastly, it supports any kind of numerical feedback over each cluster as well as any  evaluation over the entire clustering. In our tests, we will take several choice combinations of the two.

% independent  VS dependent feedbacks over clusters

% for OTA domain or not

% for OTA but not feedback clustering

\section{Problem Formulation}
\label{sec:back}
In this section, we formalize our problem.
% feedback clustering in general
Let $X$ be the set of flight searches whose elements are points in the Euclidean space $\mathbb{R}^n$, with the integer $n > 0$ representing the number of flight search features.
% 
% clustering
A {\em clustering of } $X$ { into} $k$ { clusters} is a surjective function $c:X\longrightarrow \{0,\dots, k-1\}$, where the {\em cluster} $i$ is the subset of $X$ defined by $X_i:= c^{-1}(\{i\})$, for each $i\in \{0,\dots,k-1\}$.
% In our case, the law defining $c$ depends on some algorithm whose set of inputs is denoted by $I$.
% When needed, we can highlight that $c$ depends on $I$ by writing $c_I$.
% 
%  cluster feedback
For each cluster $X_i$, we denote by $y_i\in \mathbb{R}$ the {\em (cluster) feedback}.
%  clustering feedback
For the whole clustering $c$, the {\em feedback} is denoted by the $k$-tuple $(y_0,\dots,y_{k-1})$.
An {\em evaluation of a clustering} $c$ is a real number $y$ depending on $(y_0,\dots,y_{k-1})$, denoted by $y_c=y_c(y_0,\dots,y_{k-1})$
% aim: improving clustering feedback by changing inputs in clustering phase
The problem we address is that of optimizing the evaluation $y_c(y_0,\dots,y_{k-1})$ by iteratively redefining the clustering $c$ in terms of the clustering feedback $(y_0,\dots,y_{k-1})$.

\section{SME Clustering Framework}
\label{sec:back-sme}
In this section, we recall the SME method. More details can be found in the original work~\cite{Wang2018}.
We proceed by describing the method in terms of a generic feedback evaluation.
Later on in \cref{sec:back-evaluations}, we define the feedback evaluations of our interest.

% figure of the sme

% structure of iterations

In the SME (centroid-based) clustering framework, the dataset $X$ is clusterized into clusters $X_0,\dots,X_{k-1}$ by the $k$-means algorithm initialized by $k$ random centroids $m_0,\dots,m_{k-1}$ in $\mathbb{R}^n$.
The current clustering is set to $X_0,\dots,X_{k-1}$.
% Thus, the obtained clustering $c$ depends on the set of inputs $I=\{c_0,\dots,c_{k-1}\}$. 

In the {\em evaluation phase}, the $k$-tuple $(y_0,\dots,y_{k-1})$ of cluster feedbacks is retrieved.
Moreover, the clustering is assigned its evaluation $y=y(y_0,\dots,y_{k-1})$.
% In the original work~\cite{Wang2018}, the evaluation of the first iteration is initialized to a predefined value rather than computed.

The iterative process starts with a {\em split action} over the current clustering.
In the split action, the cluster $i$ corresponding to the worst feedback value $y_i$ is selected.
Two new clusters $X_{k}, X_{k+1}$, and relative centroids $m_k,m_{k+1}$ replace the old cluster $c_i$ and centroid $m_i$ in the current status.
In addition to what authors presented in~\cite{Wang2018}, a single $k$-means iteration is performed over the current set of centroids.
This is done for technical reasons. In particular, we need to have each cluster described as the set of all points sharing the same closest centroid.
Afterwards, in the {\em merge action}, the two closest pairs of centroids among $m_1\dots,m_{i-1},m_{i+1},\dots,m_{k+1}$ are selected and their corresponding clusters discarded by the current status and replaced by their union: $X_{k+2}$ with new centroid $m_{k+2}$.
After the split and merge actions, the number of clusters is preserved.

In the {\em evolve phase}, the evaluation phase is applied to retrieve the clustering evaluation  $y_{\scalebox{0.75}{\mbox{new}}}$ relative to the current clustering.
If $y_{\scalebox{0.75}{\mbox{new}}}$ is better than $y$, then $y$ is set to  $y_{\scalebox{0.75}{\mbox{new}}}$ thus becoming the new best evaluated clustering.
In any case,  the next iteration starts with a new split action over the current clustering.
Finally, the {\em stopping criterion} is the following. The iterations are repeated till a certain value of the evaluation $y$ is reached or a certain number of iterations is performed.

% HOW to cluster: K-means for the clustering phase
\subsection{Evaluations.}
\label{sec:back-evaluations}
% K-means
% over R^n
In our study, we consider two possible evaluations of a clustering.
Both of them are aggregating internal indexes, that is numerical values associated independently over each cluster.
However, the first index we introduce is purely geometrical whereas the second one is domain-specific to flight recommendations and provides a fully external knowledge to be added in the form of feedback.
% internal index and so independency for each cluster

\subsubsection{RSS Feedback.}
\label{sec:back-rss}
% cluster rss
For each cluster $X_i$, the {\em Residual Square Sum} (RSS) is defined by
\begin{equation}
 \mbox{RSS}(i):=\frac{1}{|X_i|} \sum_{x\in X_i} {\| x - m_i  \|}^2,
 \label{eq:rss}
\end{equation}
and its evaluation over the whole clustering is obtained by taking the average weighted by cluster sizes
\begin{equation}
 \frac{1}{|X|}\sum_{i=0}^{k-1}|X_i|\cdot \mbox{RSS}(i). \label{eq:average-rss}
\end{equation}
The so-obtained evaluation is a deterministic one.

\subsubsection{Customizability feedback}
\label{sec:back-customability}
% name explained
As already stated in the introduction, customizability measures how ``learnable'' the class of customer is from a  single cluster.
In order to explain that, we need to introduce some intermediate notions.

% recommendations
First of all, we need to point out that, for each flight search, a list of 200 flight recommendation is provided by the Amadeus search engine.
The Amadeus search engine acts differently if set in terms of {\em price} or {\em value}.
% value formula
The value of a single recommended flight is obtained as a weighted combination of the flight price and other 24 not independent booking criteria such as duration, number of connections, time to wait from one flight to another, and so on.
If all weights are set to 0 then the value selection boils down to recommend flights based on price (only).
% value search
Once the weights for the value computation are set, the 200 flights are selected among several thousands according to the {\em value} associated to each single recommended flight.
%evaluation
Such a list of recommendations is evaluated in terms of {\em popularity}.
% Popularity might be computed in terms of {\em price} or {\em value}.
% popularity of recommendations
Popularity is a counter, weighted by flight ages, of the previous bookings of the same flight in history.
In these terms, a better recommendation includes more flights which have been booked a lot in the past.

% How weights are chosen
An instance of optimized weights $\bar{w}$ is found for a specific set (or a cluster) of flight searches by running an Amadeus multi-objective bayesian optimization routine.
The procedure acts on the 100 flight searches in the cluster corresponding to the most booked flights in history.
The routine provides the optimized weights $\bar{w}$.
Our assumption is that, the more the flight searches are homogeneous in terms of preferences in the booking criteria (hidden knowledge to us), the more the weights found by the Amadeus routine give recommendations with higher popularity.
Hence, the so-found weights are evaluated by picking up 100 other flight searches in the cluster randomly selected among the most booked ones.
This step makes such evaluation a non-deterministic one.
Moreover, since the absolute values of popularity might vary a lot from one set of flight searches to another, we take popularity based on price (weights set to $0$) as a stable reference.
The average popularity $pop_{\bar{w}}$ obtained over all 200 recommended flights is compared to the popularity of the recommendations obtained with weights set to 0 $pop_{0}$.
% feedback measures how good are weights for a set of flight searches
Specifically, the effectiveness of the chosen weights is measured by taking the relative change between popularity based on price and popularity based on the value obtained by the chosen weights.
% cluster customizability
\begin{equation}
 \mbox{Custom.}(i):=\frac{pop_{\bar{w}}-pop_{0}}{|pop_0|},
 \label{eq:custom}
\end{equation}
For instance, a positive cluster feedback $0.50$ means that recommendations obtained with the optimized weights are 50\% more popular than recommendations obtained with all weights set to 0, that is recommendations based on price only. However, negative cluster feedbacks can also be obtained.

% references for the method in general
% references for the method in Amadeus

To evaluate the entire clustering $X_0,\dots,X_{k-1}$, we take the following
% average customizability  formula
\begin{equation}
 \frac{1}{|X|}\sum_{i=0}^{k-1}|X_i|\cdot \mbox{Custom.}(i). \label{eq:average-custom}
\end{equation}
The so-obtained evaluation is not deterministic since values of $Custom.(i)$ might fluctuate.
We quantify later on in \Cref{sec:exp} the relevance of the fluctuation.

\section{S/M Clustering Framework: an SME variation}
\label{sec:back-s/m}
% idea: to act more locally on bad behaved clusters
% framework is slightly different from SME and EXT-SME
% - iteration: either split or merge
% - - selected according a criterion based on cluster feedback
% evolve after each split or merge call
% k is not preserved

Analogously to the SME clustering framework, in the S/M clustering framework, the dataset $X$ is clusterized into clusters $X_0,\dots,X_{k-1}$ by the $k$-means algorithm initialized by $k$ random centroids $m_0,\dots,m_{k-1}$ in $\mathbb{R}^n$.
The current clustering is set to $X_0,\dots,X_{k-1}$.
% Thus, the obtained clustering $c$ depends on the set of inputs $I=\{c_0,\dots,c_{k-1}\}$. 
\begin{figure}
\centering
\begin{subfigure}{.45\textwidth}
  \centering
  \includegraphics[width=\linewidth]{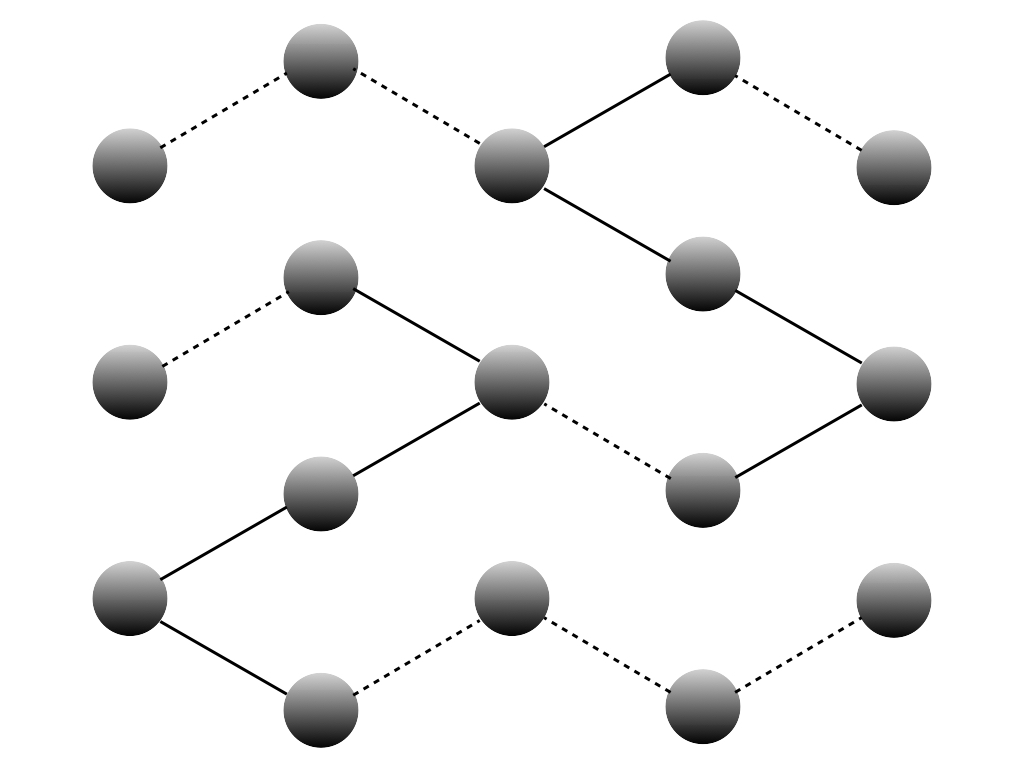}
  \caption{}
  \label{fig:pic-1}
\end{subfigure}%
\begin{subfigure}{.45\textwidth}
  \centering
  \includegraphics[width=\linewidth]{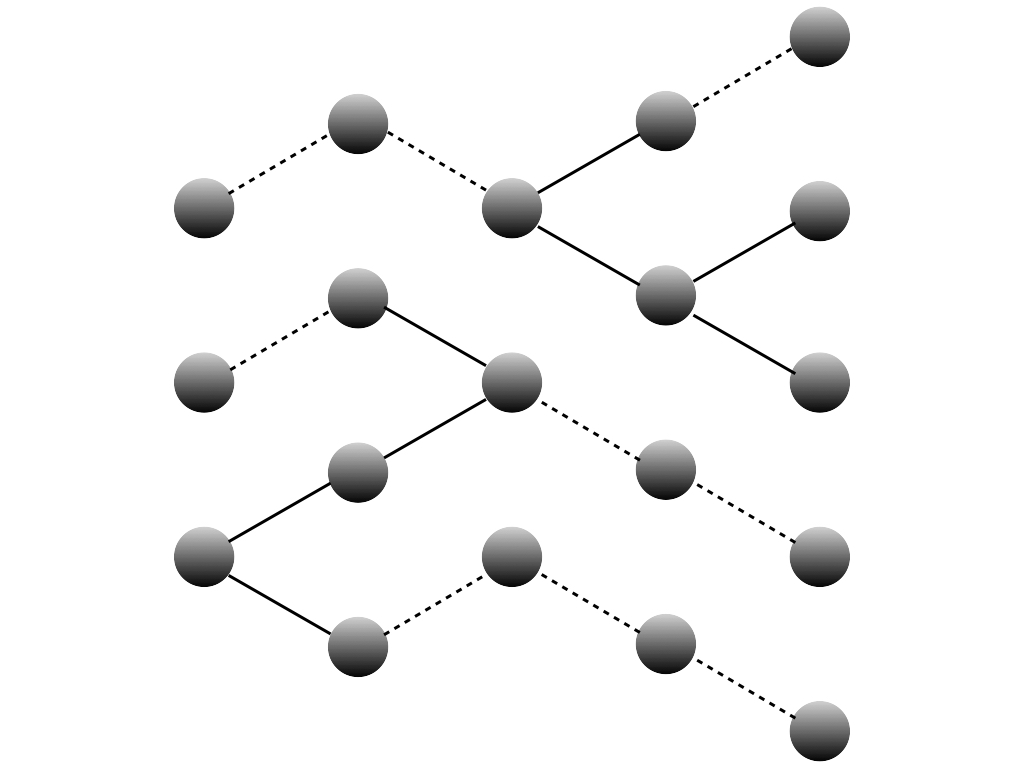}
  \caption{}
  \label{fig:pic-2}
\end{subfigure}
\caption{The logical schemas of the SME (a) and the S/M (b) methods in comparison. Each grey ball represents a cluster. On the same vertical axis, we have a clustering. Solid lines connect clusters obtained by split/merge actions from one iteration to another. Dashed lines track corresponding clusters from one iterations to another.}
\label{fig:picture}
\end{figure}

The evaluation phase corresponds to that of the SME framework.
Thus, it provides the $k$-tuple $(y_0,\dots,y_{k-1})$ of cluster feedbacks with corresponding clustering evaluation $y=y(y_0,\dots,y_{k-1})$.
% In the original work~\cite{Wang2018}, the evaluation of the first iteration is initialized to a predefined value rather than computed.

Differences are to be found in the iterative process.
This time it is decided whether to apply a {\em split action} or a {\em merge action} based on the size $s_i$ of the worst-evaluated cluster $i$.
If $s_i$ is among the first half of size values in the current clustering, then a split action is performed.
Otherwise, a merge action is performed. 
Both split and merge actions correspond to the same actions already described in \Cref{sec:back-sme} for the SME framework.

The evolve phase and the stopping criterion are analogous to the evolve phase of the SME framework.
However, as shown in \cref{fig:picture}, we underline that this time, the number of clusters $k$ is not necessarily preserved by the S/M method.
Indeed, split and merge actions can be applied a different number of times.
Moreover, in the S/M method the evaluation is performed after each split/merge action.

% % WHAT is the feedback: WO for the feebcak phase:
% \subsection{Split and Merge actions in S/M}
% \label{sec:back-s/m-criterion}

%%%% WHAT changes in the SPlit/Merge actions
% find worst cluster
% see if it is among the biggest or the smallest
% first case: apply split
% second case: apply merge
% if already done: same criteria as in SME for split or merge so that to avoid to repeat 

% How to combine feedback into clusters: independently over each cluster: by SoM

\section{Experimental Evaluation}
\label{sec:exp}
In this section, we present and discuss the results we obtained in applying feedback clustering to our use case.

% section structure
The current section is structured as follows.
In \Cref{sec:exp-settings}, we describe our test settings.
In \Cref{sec:exp-sme}, we test the impact of methods according to the kind of feedback they are acting on.
In \Cref{sec:exp-all-methods}, we compare all tested methods to the specific task of increasing customizability feedback, whatever their feedback.

\subsection{Test settings}
\label{sec:exp-settings}
%methods and labels
In this section, we outline the settings considered for our tests.
The methods to be tested are the SME method introduced in \Cref{sec:back-sme} and the S/M method introduced in \Cref{sec:back-s/m}.
Each method, can be implemented with respect to the RSS or the customizability (Custom.) feedback introduced in \Cref{sec:back-evaluations}.
Hence, we have four methods to be tested:
\begin{itemize}
 \item {\em SME(RSS)}: split-merge-evolve method first introduced in~\cite{Wang2018} (see \Cref{sec:back-sme}) where the RSS index evaluates each cluster. Each SME iteration is performed over the worst-valued cluster (maximum under RSS)
 \item {\em SME(Custom.)}: split-merge-evolve method first introduced in~\cite{Wang2018} (see \Cref{sec:back-sme}) where the customizability index evaluates each cluster. Each SME iteration is performed over the worst-valued cluster (minimum under customizability)
 \item {\em S/M(RSS)}: split-merge-evolve method is modified so that split and merge phases are separated (see \Cref{sec:back-s/m}). The worst-values cluster (maximum under RSS) is either split or merged to its closest cluster according to its size. 
 \item {\em S/M(Custom.)}: split merge evolve method is modified so that split and merge phases are separated (see \Cref{sec:back-s/m}). The worst-values cluster (minimum under customizability) is either split or merged to its closest cluster according to its size.
\end{itemize}

%domain: flight search representations
Flight searches are represented as points in the 8-dimensional space.
Indeed, along with origin and destination not included in the 8 dimensions, each flight search provides: distance between origin and destination, advance purchase, stay duration, number of passengers, number of children, geography (categorical value to distinguish among domestic, continental and intercontinental flight searches), departure day of the week, return day of the week (taking values from $0$ to $6$).
Thus, we have heterogeneous features mixing numerical and categorical values.
Moreover, our features might be dependent from one another.
Every feature value is numerically treated as a real number value.
For technical simplicity reasons, only round-trip searches are considered.
Beyond the 8 features, each flight search comes associated with its own origin/destination data, that is the departure and arrival airport, respectively.
Origin and destination are not directly used in our clustering phase.

% datasets
In our tests, we considered 3 datasets varying according to the country of origin (see \ref{tab:datasets}).

\begin{table}[h!]
\centering
\begin{tabular}{@{}llc@{}}
Name  & Size  &  Expected Relative Change          \\ \midrule 
% US    & 4.7 M &   0.126 & 0.236 \\
FR    & 348 k 
% & 0.022 
& 0.197\\
GB    & 4 M  
% & 0.056  
& 0.201\\
AR/BR & 1.5 M 
% & 0.041 
& 0.206
\end{tabular}
\caption{Datasets characteristics.}
\label{tab:datasets}
\end{table}

Moreover, datasets are chosen to be various in terms of size and expected relative change in their customizability evaluation.
Indeed, as already mentioned, customizability evaluations are subject to value fluctuation. 
We quantify it over each dataset. For each number of clusters $k$ in $[2,3,4,5,6,7]$ we measure the expected relative change over $10$ customizability evaluation calls over the same clustering of $k$ clusters.

% settings
For each dataset, methods are called with a number of initial clusters $k$ varying in $[2,3,4,5,6,7]$.
The SME(RSS) and the SME(Custom.) methods are tested over all datasets by applying 6 iterations.
The S/M(RSS) and the S/M(Custom.) methods are tested over all datasets by applying 12 iterations to have the same number of elementary operators as for the SME methods.
Independently from the method, both evaluations RSS and customizability are stored.
%%% figure example iterations over the same dataset and same initial clusters to cooment similarities and dissimilarities
%%% ....
%% Customizability and RSS have independent behavior: this confirms relevance of feedback clustering
%% S/M: the number of clusters is generally increasing 
%%% rules can be changed to rebalance that
% we remark that both SME and SoM do not improve the clustering quality at each feedbakc iteration

% aim
Our aim is to measure the impact of the tested methods on the initial clustering evaluation: $RSS_0$ for the RSS feedback and $Custom_0$ for customizability.
The initial evaluation is compared to the best obtained during the method call: $\min RSS$ for the RSS feedback and $\max Custom$ for the customizability feedback.
The comparison is numerically obtained by taking the relative change in between initial and best iteration. 

% machine specs
% timings

\subsection{Test 1: own feedback impact evaluation of  SME}
\label{sec:exp-sme}
\begin{figure}
% \caption{Some Typical Commands}
\centering
\includegraphics[width=.6\textwidth]{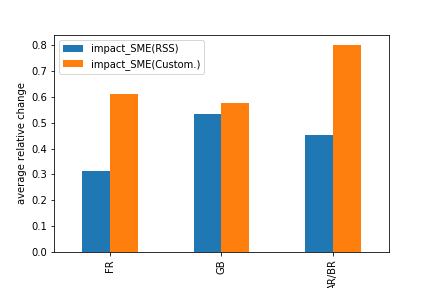}
\caption{Average relative change by dataset in between initial and best iteration under the SME method. {\em  impact\_SME(RSS)} and {\em  impact\_SME(Custom.)} reports relative change under the RSS evaluation and customizability evaluation, respectively.}
\label{fig:impact-sme}
\end{figure}

The first comparison we report is that in between the SME(RSS) method and the SME(Custom.).
The impact of each method for a given number $k$ of initial clusters is taken with respect to the feedback evaluation leading the process. Indeed, for SME(RSS), the impact (blue bars in  \Cref{fig:impact-sme}) is the relative change between initial and best RSS evaluations along the process.
The average of all impacts is taken over runs with $k$ varying in $\{2,3,4,5,6,7\}$.
Analogously, the impact for SME(Custom.) (orange bars in  \Cref{fig:impact-sme}) is the relative change between initial and best customizability evaluations.

% SME's comparison results/discussion
We can use the SME(RSS) impact as a reference for SME(Custom.).
For all datasets, the SME(Custom.) presents higher values. 
We believe that this is partially due to the fact that the RSS feedback is somehow already optimized by the $k$-means algorithm whereas the customizability feedback is a purely external feedback.
The order of magnitude of the two impacts is the same.
This, in lack of a ground-truth to measure the effectiveness of the SME(Custom.), suggests reliability on the impact evaluations.
For AR/BR and FR datasets, values of SME(RSS) are doubled by those of SME(Custom.).
For GB dataset, values are closer.

For the RSS evaluation (blue bars), our measures confirm the effectiveness of the SME method applied to the domain of online flight searches.
% Differently from what done in~\cite{Wang2018} to evaluate the SME method over synthetic data, we considered only an internal index (the RSS index) to measure the clustering quality.
%
%%% to be added: ARI comparison
%
For customizability (orange bars), we found all dataset impacts more relevant compared to the dataset expected relative change reported in Table~\ref{tab:datasets}.
This confirms the adaptability of the SME method being not limited to standard clustering quality indexes (RSS), rather for improving a domain specific quality index such as customizability.

%% RSS confirms its effectiveness: rss always improved (relatively)
%% RSS is deterministic allows us to observe wether or not Customizability being not deterministic impacts too much:  it should be better investigated but not too much in this use case

\subsection{Test 2: SME and S/M for segmenting online flight searches}
\label{sec:exp-all-methods}
%% test the effectiveness of S/M over SME 
Similarly to what done in \Cref{sec:exp-sme}, we compare methods {\em SME(RSS)}, {\em SME(Custom.)}, {\em S/M(RSS)}, and {\em S/M(Custom.)}.
This time, each method is evaluated with respect to customizability feedback.
This means that methods led by customizability feedback are evaluated as in Test 1.
Instead, methods led by RSS evaluation are evaluated by considering as their best clustering that one obtained under RSS.
The corresponding customizability evaluation is stored as the reference one.
Then, the relative change is computed with respect to initial and referenced customizability evaluations.
Afterwards, the average among all possible initial clusters is computed.

\begin{figure*}
% \caption{Some Typical Commands}
\centering
  \label{fig:impact}
  \includegraphics[width=.6\textwidth]{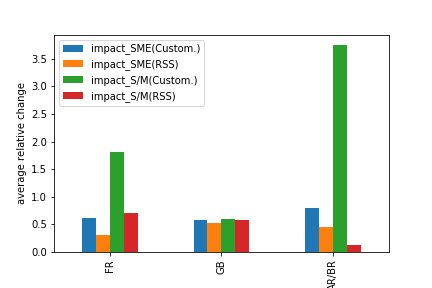}
  \caption{Method impacts measured by average relative change of customizability.
  Results are shown by dataset.}
\end{figure*}

% \usepackage{booktabs}
% \usepackage{multirow}
\begin{table}[h]
\centering
\begin{tabular}{@{}lllllll@{}}
\multirow{2}{*}{Dataset} & \multicolumn{6}{c}{Number of Initial Clusters}                                                                                                \\ \cmidrule(l){2-7} 
                         & \multicolumn{1}{c}{2} & \multicolumn{1}{c}{3} & \multicolumn{1}{c}{4} & \multicolumn{1}{c}{5} & \multicolumn{1}{c}{6} & \multicolumn{1}{c}{7} \\ \midrule
FR                       & 0.072                 & 0.208                 & 0.208                 & 0.180                 & 0.221                 & 0.251                 \\
GB                       & 0.565                 & 0.360                 & 0.379                 & 0.466                 & 0.447                 & 0.357                 \\
AR/BR                    & 0.183                 & 0.141                 & 0.264                 & 0.257                 & 0.342                 & 0.272                
\end{tabular}
\caption{Average initial customizability evaluations by number of initial clusters}
\label{tab:initial-custom}
\end{table}

%% Customizability more effective than RSS to improve Customizability
As expected, we found methods {\em SME(Custom.)} and {\em S/M(Custom.)} to be more effective than the others since they are led by the feedback we want to optimize.
%% S/M most of times has the best improvement from initial to best clustering
In particular, method {\em S/M(Custom.)} outperforms the others while {\em SME(Custom.)} has results more similar to the methods driven by RSS.
%% fluctuation
Compared to dataset expected relative changes in Table~\ref{tab:datasets}, we register impacts higher than expected in all cases but for {\em S/M(RSS)} over dataset AR/BR.

\begin{figure}
\centering
\begin{subfigure}{.45\textwidth}
  \centering
  \includegraphics[width=\linewidth]{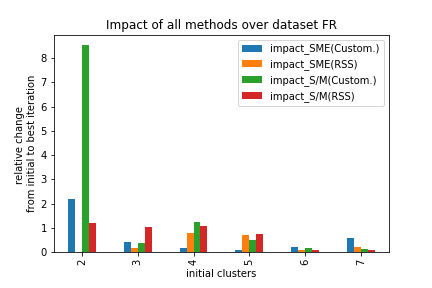}
%   \caption{Customizability by datasets}
  \label{fig:FR-custom}
\end{subfigure}%
\begin{subfigure}{.45\textwidth}
  \centering
  \includegraphics[width=\linewidth]{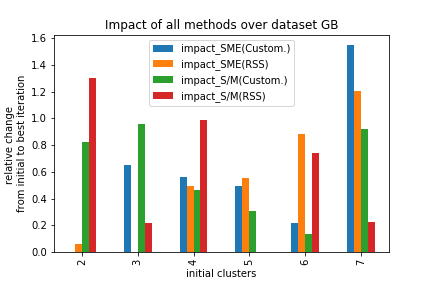}
%   \caption{RSS by datasets}
  \label{fig:GB-custom}
\end{subfigure}
\begin{subfigure}{.45\textwidth}
  \centering
  \includegraphics[width=\linewidth]{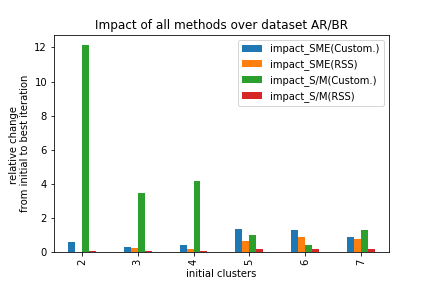}
%   \caption{Customizability by datasets}
  \label{fig:AR-BR-custom}
\end{subfigure}%

\caption{Customizability relative changes by initial number of clusters $k$}
\label{fig:custom-by-k}
\end{figure}

%%% the 4 datasets have very different RSS values (cluster quality is various)
In Figure~\ref{fig:custom-by-k}, we show our results for each dataset as the number of initial clusters $k$ varies.
This detailed view allows us to notice  a high variance in terms of both datasets and number of clusters $k$ within the same dataset.
Indeed, dataset GB presents very low relative changes compared to the others probably due to higher initial evaluations as shown in Table~\ref{tab:initial-custom}.

% S/M generally better
We see that most of peaks corresponds to S/M(Custom.) method. 
However, the GB dataset shows a more various behavior and its best relative change is obtained by SME(Custom.) over $7$ clusters.
Remarkably for $k=6$, we have the two $RSS$-driven methods getting better results than the customizability-driven ones.
The results in Figure~\ref{fig:custom-by-k} might be helpful to determine the number of optimal clusters $k$ to subdivide a given dataset.
However, we recall that for methods {\em S/M}, the setting of $k=2$ does not mean that the best found clustering contains $2$ clusters.
For instance, for dataset GB with initial clusters $k=2$,  the best clustering contains $5$ clusters although the process started with $2$. It means that, in that case, the split phase was called more times than the merge one.
The S/M(Custom.) bar for $k=2$ is significantly higher that all bars at $k=5$, in particular those for the SME method.
This suggests that, in that particular case, we were able to find a better segmentation into $5$ parts by starting with $2$  clusters than those obtained by recombining $5$ clusters.
This is one of the strength points in the S/M method over the SME one.
% AR/BR green bars
For dataset AR/BR, we observe remarkable peaks for $k=2,4$ and method S/M(Custom.).
The corresponding best found clusterings contain $6$ clusters each where a particular cluster has been well-evaluated.
The fact the the bar for $k=2$ is higher that $k=4$ is due to the initial evaluation being different (see Table~\ref{tab:initial-custom}). 
This suggests a more local behavior of the S/M method compared to SME in finding particularly well-behaved clusters, whereas the SME method seems to reward the overall improvement.

\section{Conclusions}\label{sec:conclusions}
In this paper, we have presented how feedback clustering techniques can be applied to a specific use case: recommendations for online flight searches.

% % %  our feedback clustering framework has independent feedback cluster evaluations
% we have discussed how to assess our case study in the framework of feedback clustering
In particular, we have shown how our use case can be formalized within the feedback clustering framework.
This has been achieved by considering two cluster evaluations as our feedbacks independently taken for each cluster.
The former is the RSS index, that is a well-known deterministic cluster index measuring how points within a cluster are spread apart from the cluster centroid in average.
The latter has been introduced in this work and called customizability.
Customizability is a non-deterministic value which is domain-specific to flight searches. 
Indeed, customizability measures how much a specific optimization process we exploit in our use case is efficient over some cluster.
The optimization process is meant to assign a specific customer behavior (schematically: business, leisure traveller, family group, etc.) to some cluster by learning it from flight searches only.
Customizability measures how satisfying is the customer behavior detected by the optimization process in terms of efficiency of the corresponding flight recommendations found.

We detected the Split-Merge-Evolve (SME) method introduced in~\cite{Wang2018} as a suitable one to be tested.
Indeed, it is very flexible in terms of the core clustering algorithm in use (here it is the $k$-mean algorithm) and it acts on any numerical feedback being independent over each cluster (required by our use case).
Based on that, we defined a clustering evaluation depending on each cluster feedback as the average cluster feedback evaluations weighted with cluster sizes.
% The SME method consists in iteratively modifying some clustering obtained through the $k$-means algorithm.
% Each iteration produces a new clustering into $k$ parts after splitting the worst cluster and merging the two closest ones. Then, the new clustering is evaluated again.

% % % SME confirms its effectiveness in the domain of online flight searches
% % % ADDRESSING effectiveness of a domain-specific feedback
% non-deterministic
% no ground-truth and best clustering criteria not correlated to standard clustering quality measures
% customizability independent from RSS, thus more informative
% % % domain-specific feedback has better results than standard cluster quality feedbacks
%
Our first contribution has been to compare the effectiveness of a domain-specific feedback (customizability) over a standard cluster index (RSS).
In order to do so, we first tested the SME method implemented with the RSS and the customizability feedbacks over $3$ heterogeneous datasets of flight searches.
We measured how much the SME method improves the global clustering feedback, depending on the chosen feedback.
Results in \Cref{sec:exp-sme} confirm effectiveness of the SME method for our domain for both feedback choices.
For the RSS feedback, results confirmed the efficiency of the SME method already obtained in~\cite{Wang2018} for the case of synthetic data.
For the customizability feedback, the impact is higher than for RSS.
In our opinion, this is partially due to the customizability feedback being purely external whereas the RSS index improvement is already part of the $k$-mean algorithm's target.
% Nonetheless, in our use case, it has been proper to introduce a domain-specific index.
This provides a real-world use case where the feedback clustering framework has been successful.
However, the two impacts had the same order of magnitude and this confirms the flexibility of the SME method under feedback evaluation changes.
Moreover, this, in lack of a solid clustering ground-truth for a domain-specific feedback, strengthen the reliability our results. 

Secondly, in \Cref{sec:exp-all-methods}, we evaluated SME method under the RSS and the customizability index in improving customizability, specifically.
As expected, we found in average better results with the customizability feedback.
However, for particular choices in the number of initial clusters per dataset ($5$ out of $18$), our results presented better improvement under the RSS instead of the customizability feedback. 
In general, our results provide an instance of effectiveness of a domain-specific feedback over a standard cluster index.

% % % introduction of a variation on the SME method : S/M
% S/M can be applied to any domain in theory
As a second contribution, we introduced a variation of the SME method called S/M and tested it.
S/M is theoretically as flexible as SME in terms of cluster feedback to drive it.
The S/M method differs from SME only in the way clusterings are altered from one iteration to another.
Specifically, the worst cluster under the feedback evaluation is either split or merged to the closest one according to its size in terms of number of points.
In \Cref{sec:exp-all-methods}, we compared the customizability improvement obtained under the S/M method to the SME method.
Our results  showed that S/M behave like SME in presenting better results when driven by customizability feedback rather than by RSS.
As for the comparison in between S/M and SME, the former had average better results over all datasets with a remarkable gap for $2$ datasets out of $3$.
We suggest that this behavior is probably due to S/M acting more locally than SME.
Indeed, on our domain, the combination of split and merge operators in S/M seems more suitable to isolate clusters with bad feedbacks or to favoring clusters with very good scores.
Other domains might prefer the SME method where this polarized effect is tamed.\\

%%% futures
Ongoing work directions include the followings.
First, we are setting further tests to compare possible clustering evaluations based on cluster customizability other than taking the average weighted on size.
This would help us in rewarding single cluster good peaks.
Secondly, we are designing new split and merge combinations so that to better handle the fluctuation of customizability values for each cluster. For instance, we could consider multiple runs of split and merge operators at each iteration.
Lastly, we are working on comparing the impact on flight recommendations of SME and S/M to other frameworks such as ensemble clustering~\cite{SujoySimon2020arXiv}. This would provide further tests for the effectiveness of customizability index, specifically and as a domain-specific feedback.
\bibliographystyle{splncs04}
\bibliography{ismis-ref}

\end{document}